# Separating the influence of Brain Signals from the Dynamics of Heart

P G Vaidya
National Institute of Advanced Studies, Indian Institute of Science Campus, Bangalore 560012, India, Email: pgvaidya@nias.iisc.ernet.in

## ABSTRACT

ECG signals appear to be quite complex. In this paper, we present results, which show that a normal ECG signal, which is a function of time can be transformed into a relatively simpler signal by stretching the time in a predetermined way. Before such a transformation, if you were to analyze various packets of the data for the Trans-Spectral Coherence (TSC) you could confirm that the signal indeed is very complicated. This is because the TSC gives us an idea of how various harmonics in a spectrum are related to one another. The coherence dramatically improved once we found an intermediate variable.. However, there was one hurdle. To find this variable, we needed to postulate that the complexified data lies on a 4-sheeted Riemann-Surface. With this insight, we could identify a proper time transformation which led to extremely high TSC.

The transformed signal is quite simple. We can now rearrange the ECG data in terms of a set of functions in an affine space, which we can explicitly calculate from the data. This reveals that the dynamics of the heart, if freed from the external influence is quite simple.

# Introduction

ECG signals appear to be quite complex. We can speculate that this complexity is due to the coupling of the whole body with heart. Cardiac cells are capable of autonomous oscillations. If the analogy and simulations with simple Vander-Pol like oscillator are correct, an external impulse pushes the oscillation outside of its limit cycle. If the limit cycle were to be stable, for small impulses the system returns to the limit cycle, with a possible re-adjustment of phase. Our goal in this paper is to isolate the effect of this phase change. The paper presents the results, which show that normal ECG signals, which are a function of time, can be transformed into a relatively simple signal by stretching the time in a predetermined way.

A method called TSC(Trans-spectral coherence ) played an important role in this work. This is similar to phase coherence. It consists of taking various subsections (windows) of a data set and doing a DFT on each of these.

To take a simple case, supposing in a given window we find the component at $\omega$ has a phase $\theta$ and the component at 3 $\omega$ has a phase $\Phi$ then we compute:

$$\tau = 3\theta - \Phi$$

If this $\tau$ is a constant from window to window, we say that the TSC, for 1 to 3, is unity. Specially, we compute the sum of square of the averages of the cosines and sines of tau in each window.

This quantity is one for prefect coherence and it is a “translation invariant” for each window. It is zero if there is no coherence at all. More advanced versions of TSC are in Reference 1 and 2.This is because the TSC gives us an idea of how various harmonics in a spectrum are related to one another. For a system governed by a low order dynamics (even if chaotic) it has been found and recently theoretically validated that the TSC would be quite high. Coming back to ECG data analysis, if you were to analyze various packets of the data for the Trans-Spectral Coherence (TSC) you could confirm that the signal indeed is very complicated

Since noise reduces TSC, as a first step, a very efficient filtering was carried out. Even then the results were quite poor. They dramatically improved once we transformed time to account for the external influences. To do this, we first tried the Hilbert transform, since it gives a phase as natural output. This was not successful. Then we used the complex data sets in the accompanying paper. This worked, but there was still one remaining hurdle. The phase of the Complexification itself would not lead to packets with good TSC. It was then postulated that the complexified data lies on a 4-sheeted Riemann-Surface. With this insight, we could identify a proper transformation which led to extremely high TSC.

## ECG Data provided by Dr Pradhan of NIMHANS

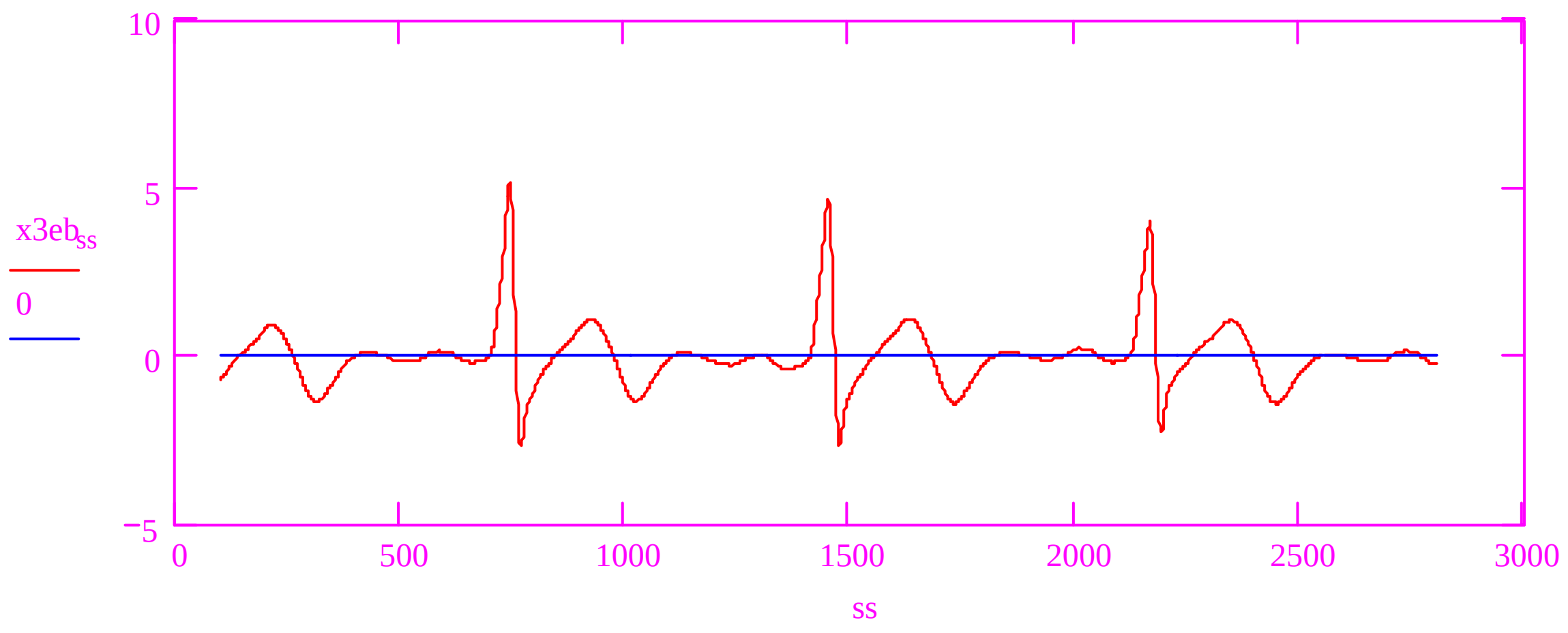

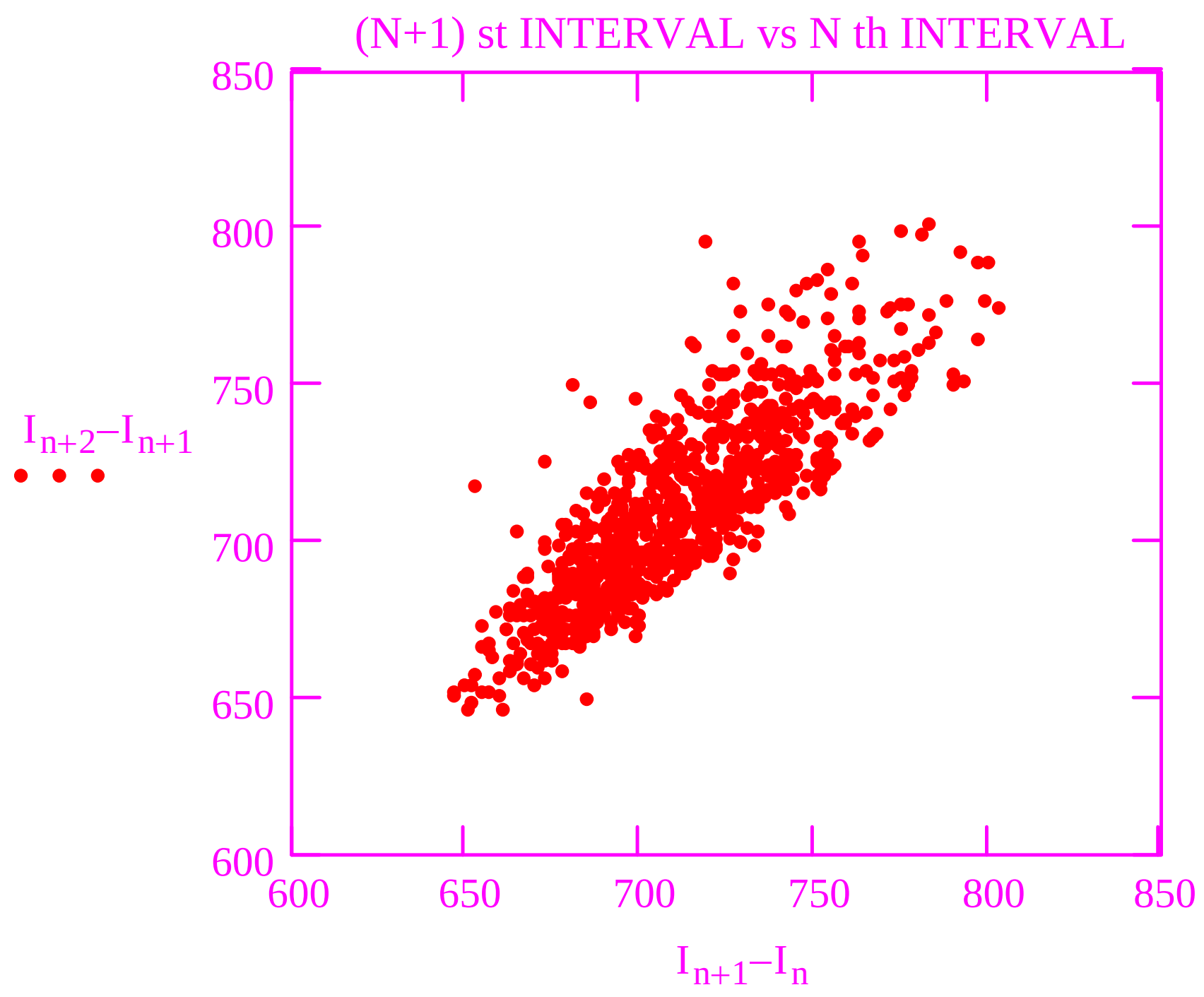

RR Interval Plot

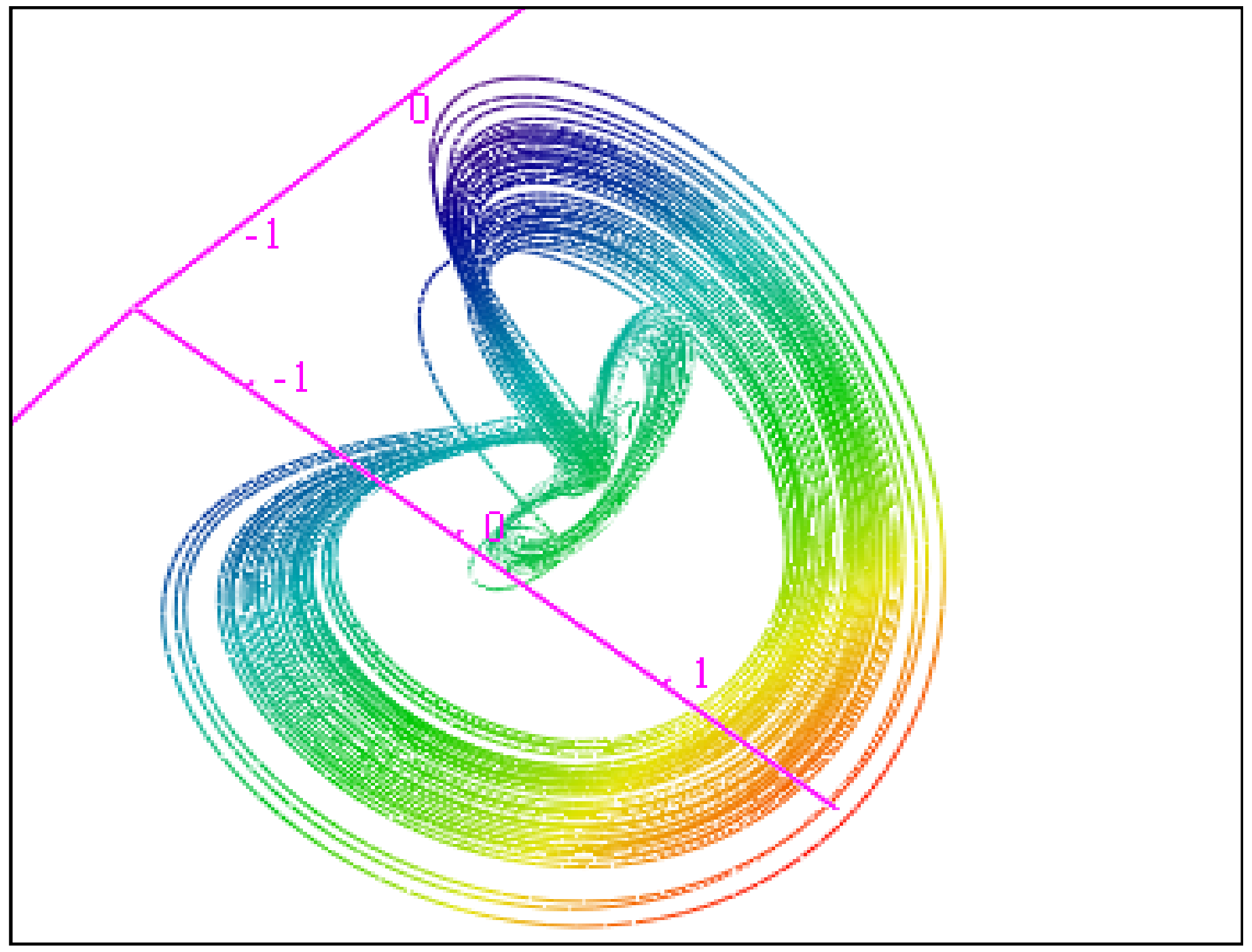

( X5low , Y5hi , X5hi )

**Phase portraits generated by embedding vectors**

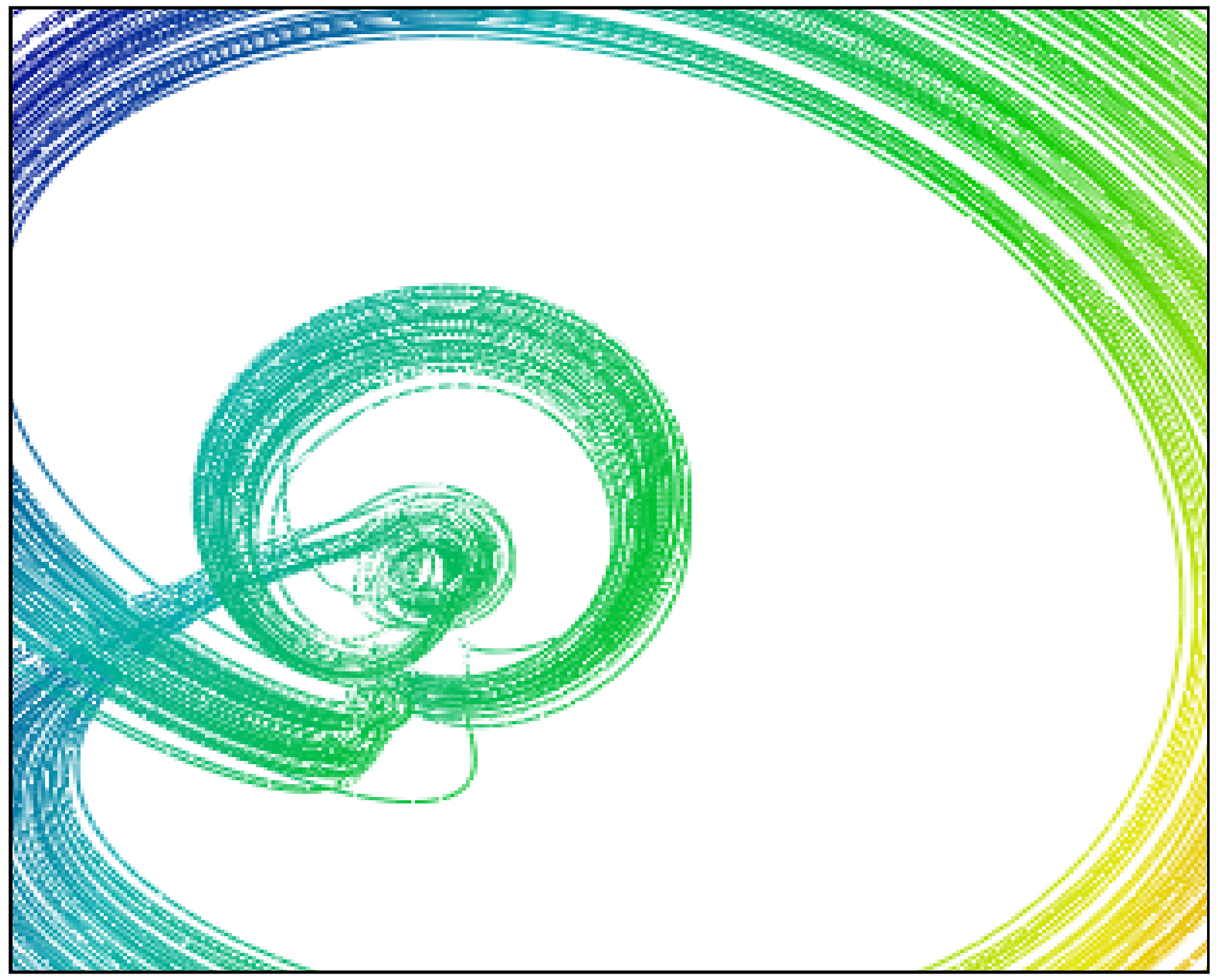

( X5low , Y5hi , X5hi )

# Trans-Spectral Coherence & Riemann Sheet and its Unfolding

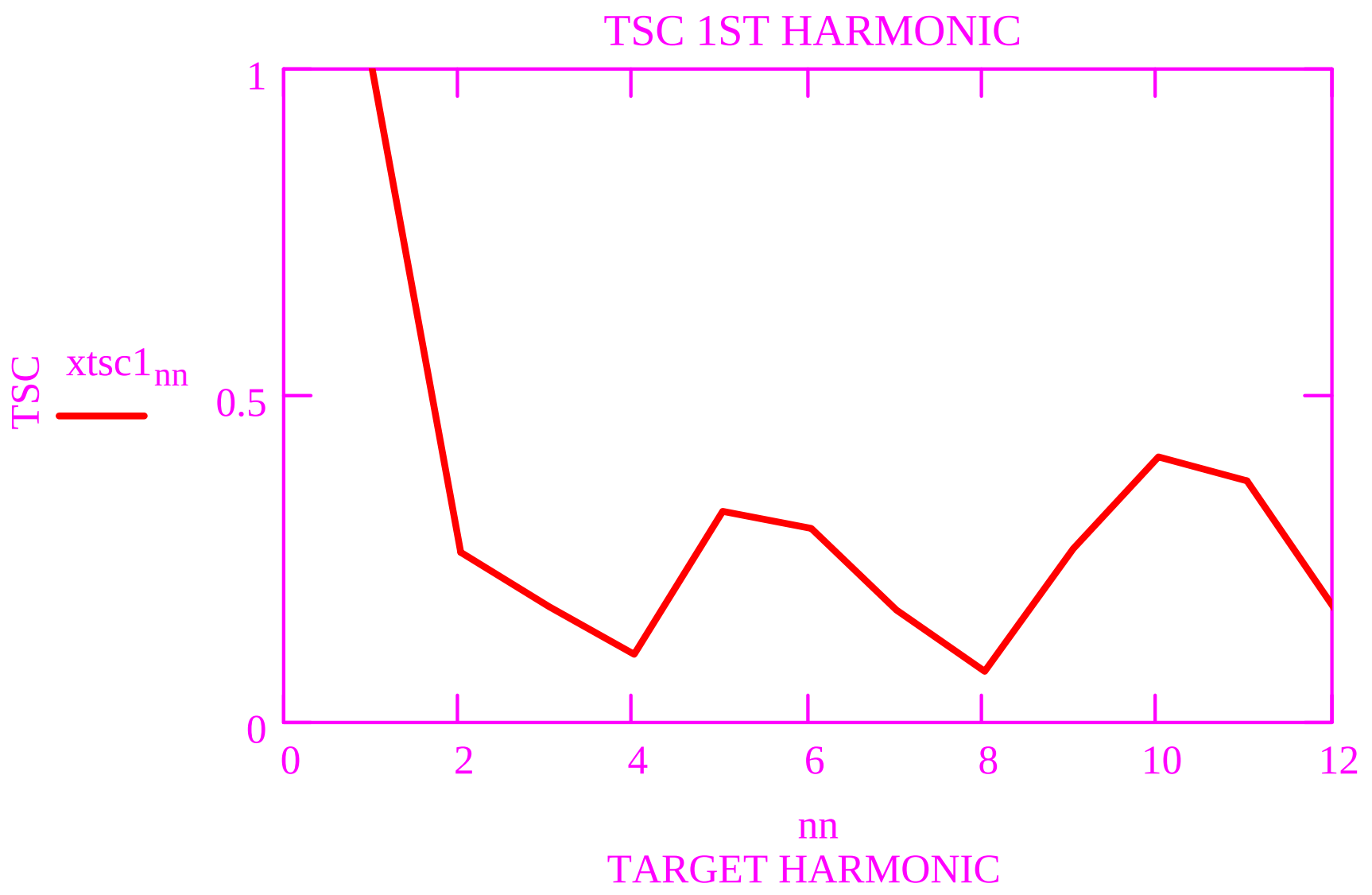

**TSC of the first harmonic is very poor (other than the trivial 1 to 1 case which has to be unity)**

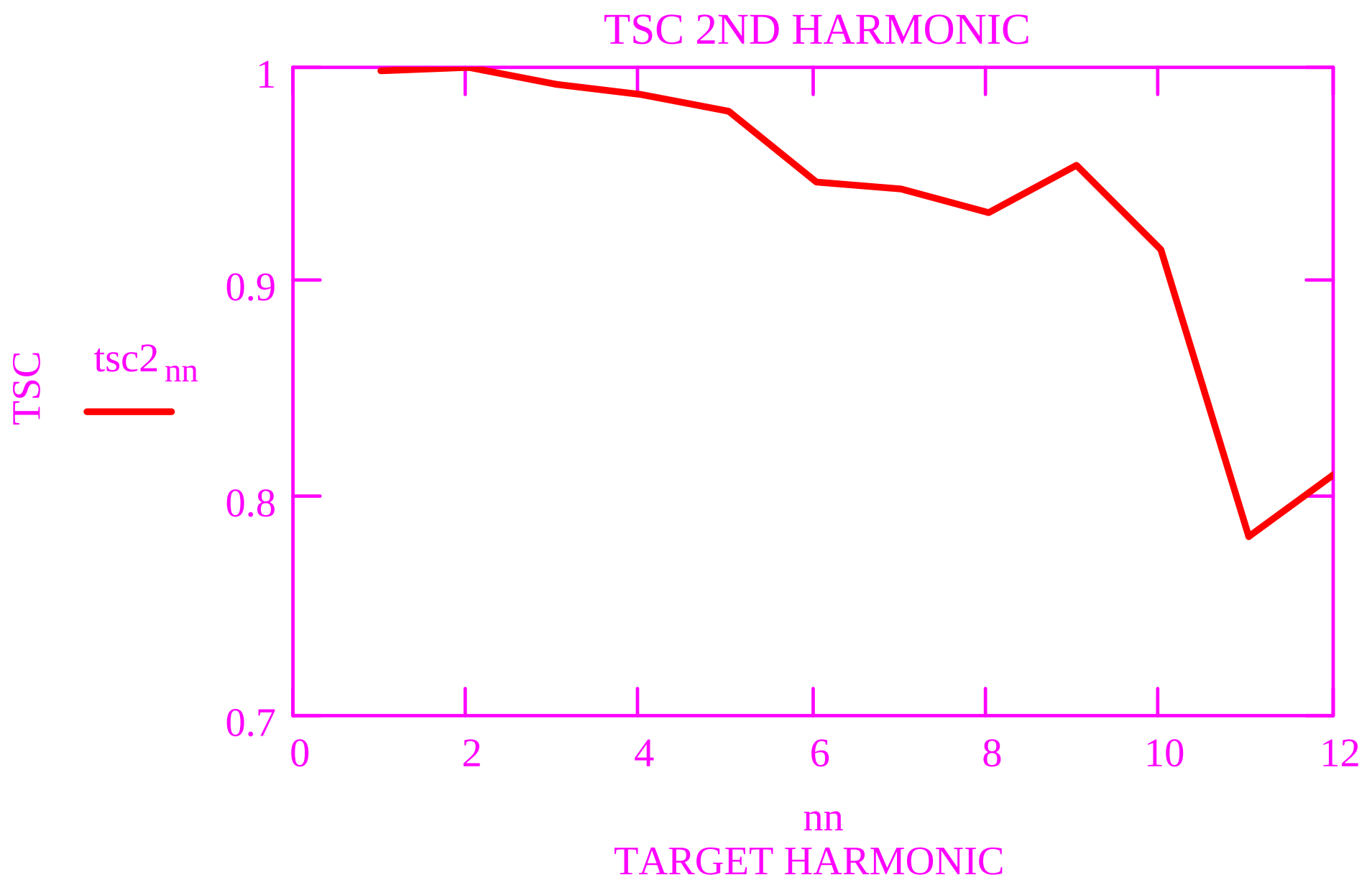

**Second harmonic is still better**

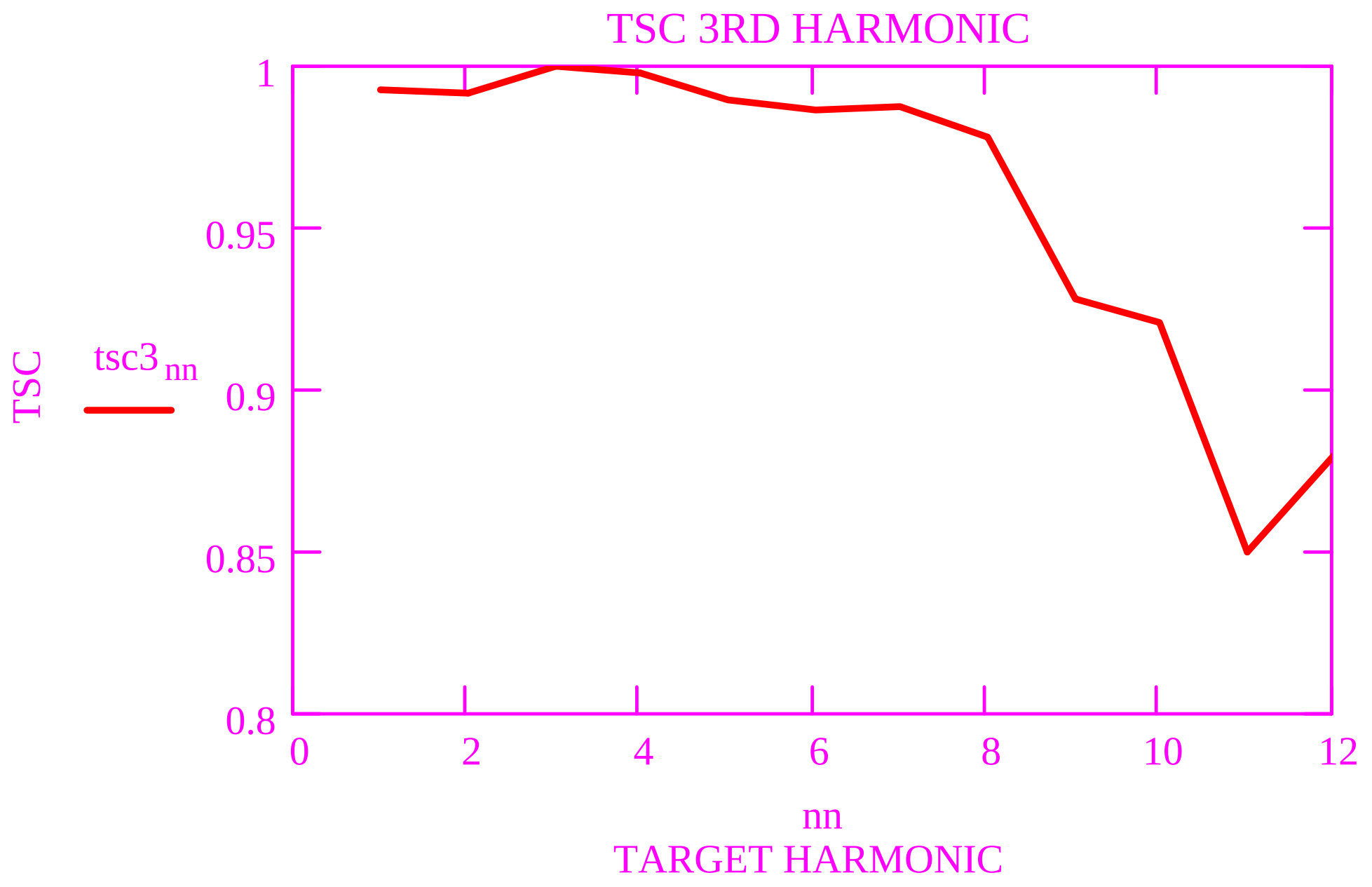

**Third harmonic is still better**

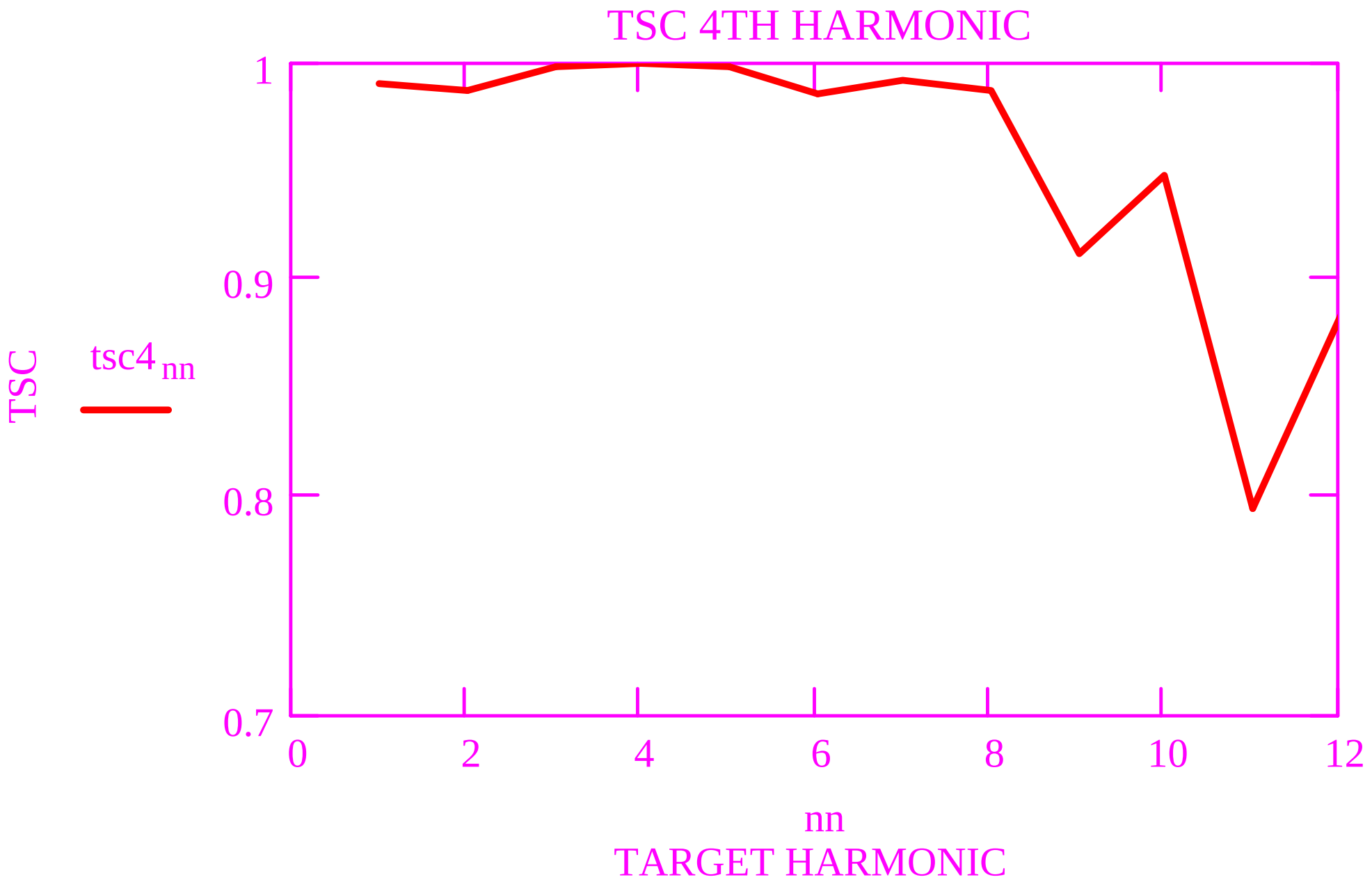

**Best so far (giving a hint of the further results**

**Plot of phase Vs. the signals showing a four cycle**

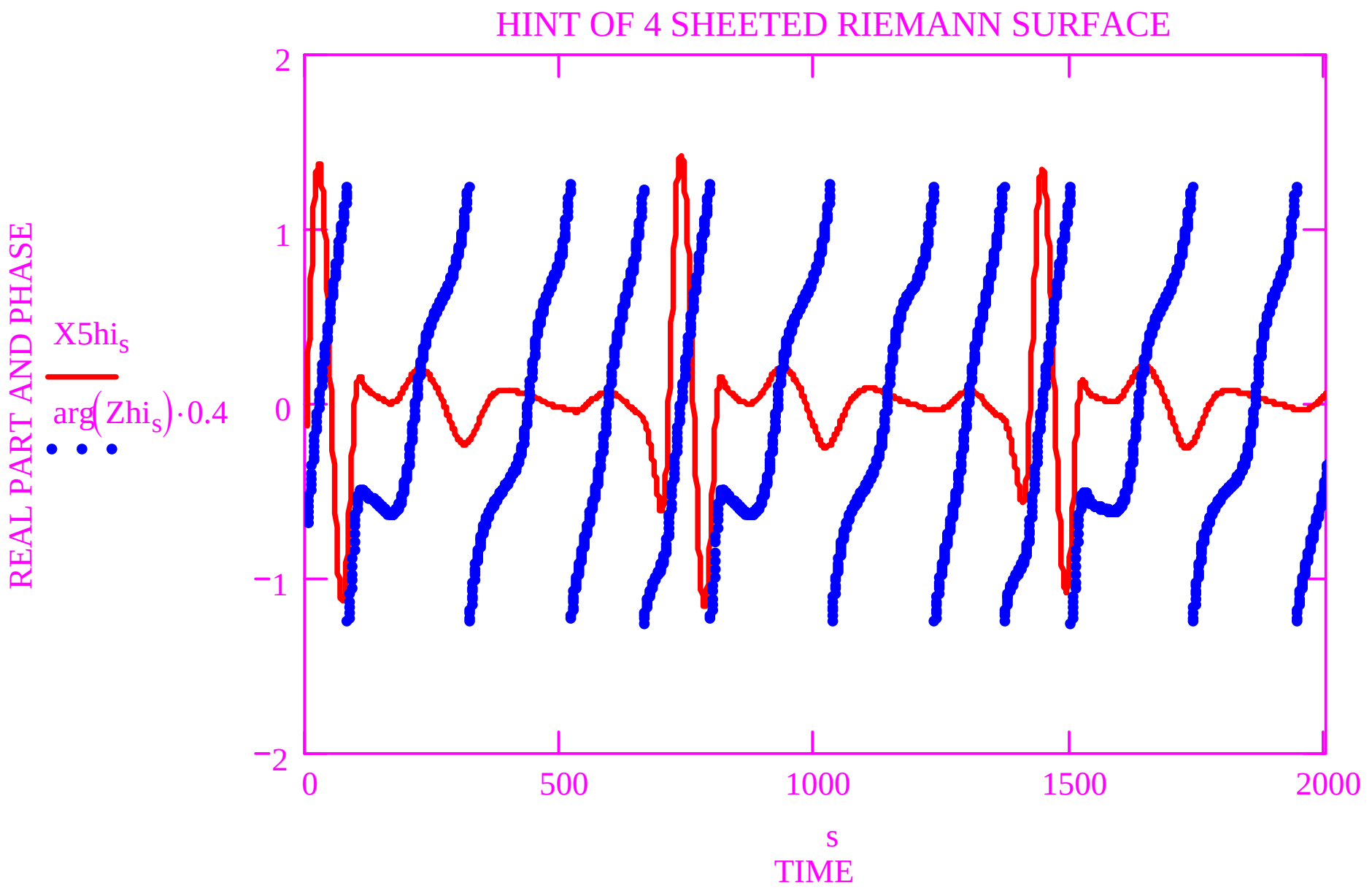

**The plot of the signal in the complex plane shows many intersections. However, for the special filtering and this specific data we can carry out a transformation:**

$$W = |Z| e^{\left(\frac{i.\arg ument(z)}{4}\right)}$$

**This has an effect of unfolding a 4 sheeted Riemann sheet so that the self intersection are eliminated.**

# Unfolded Phase Plot & State Space picture in 3D

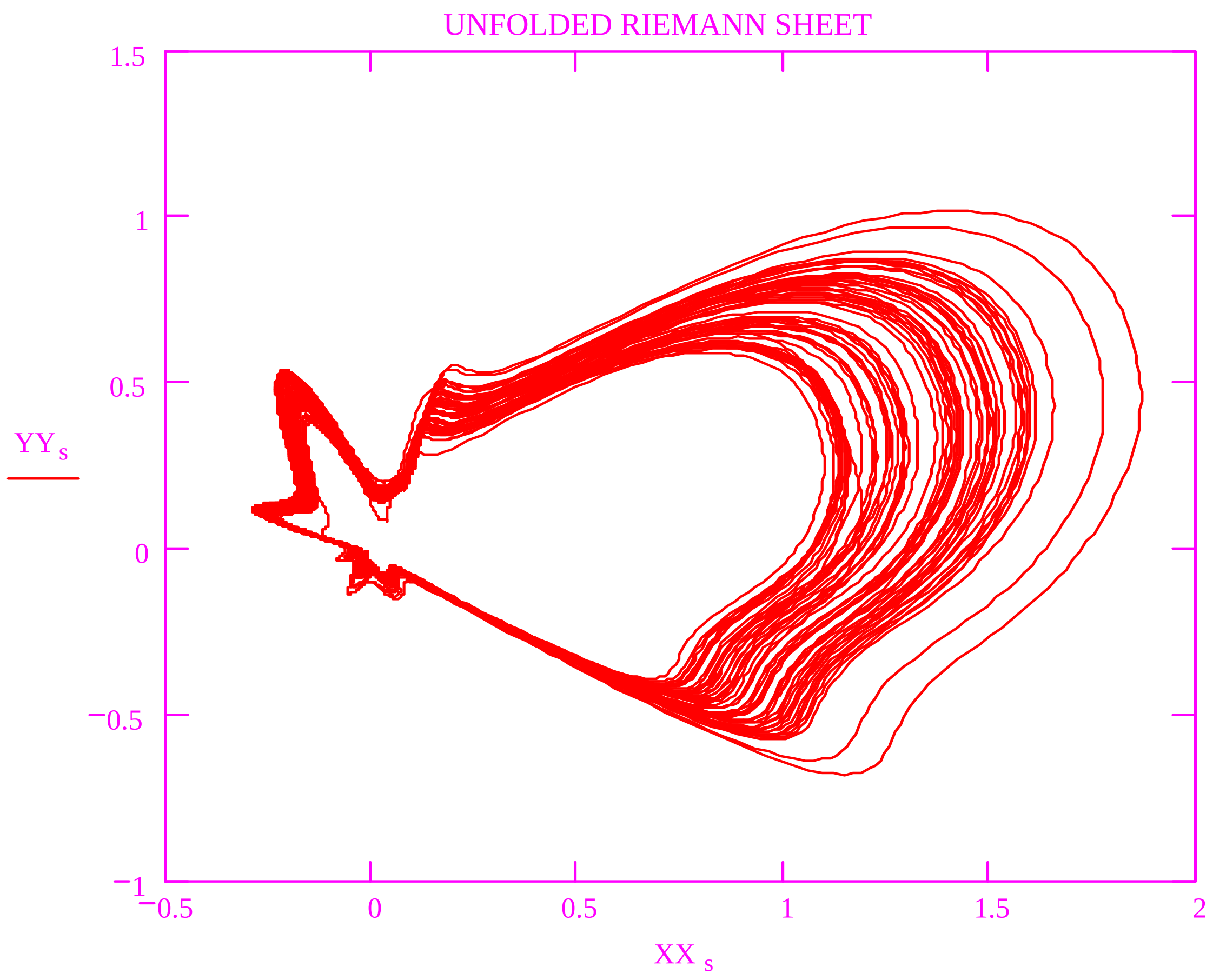

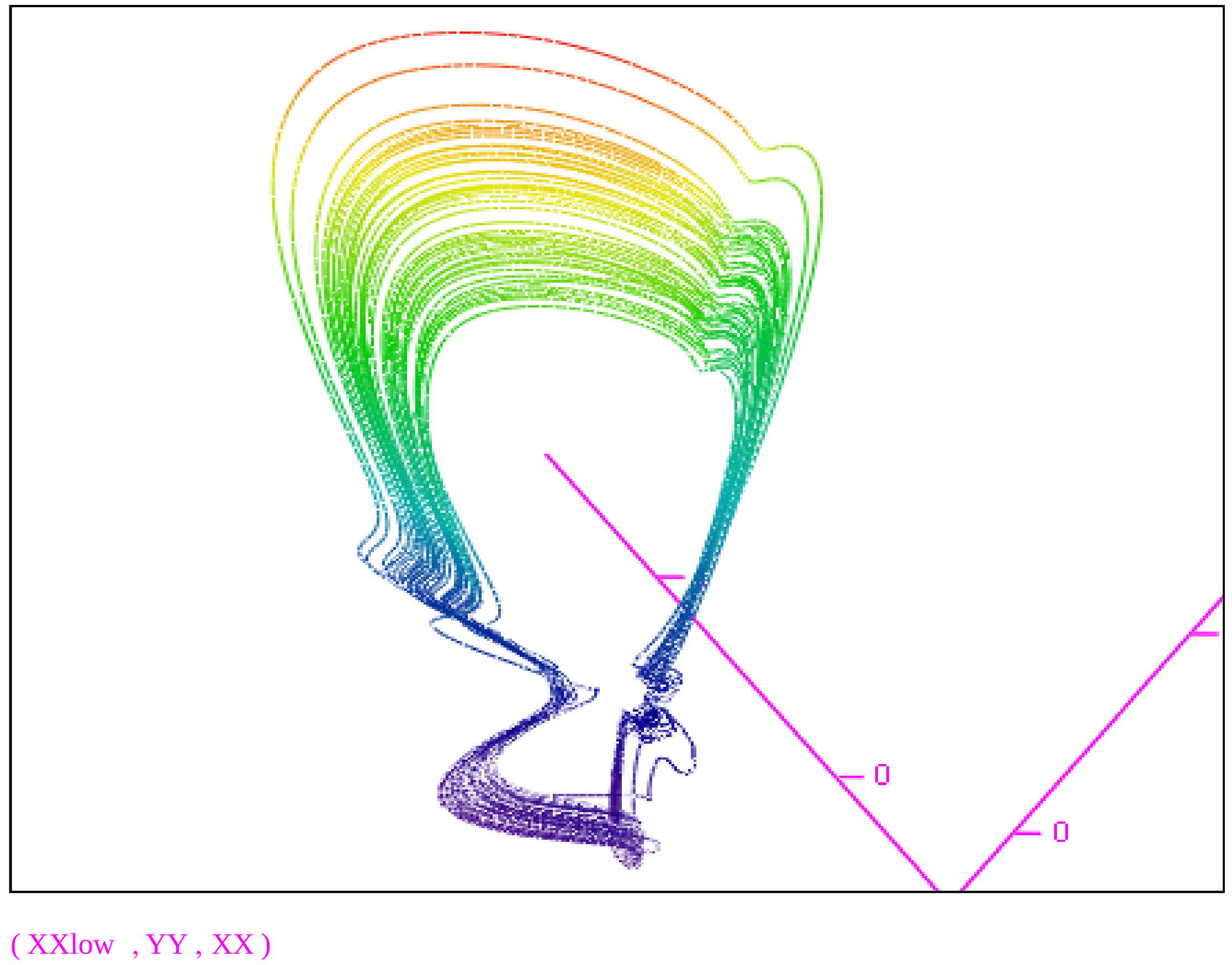

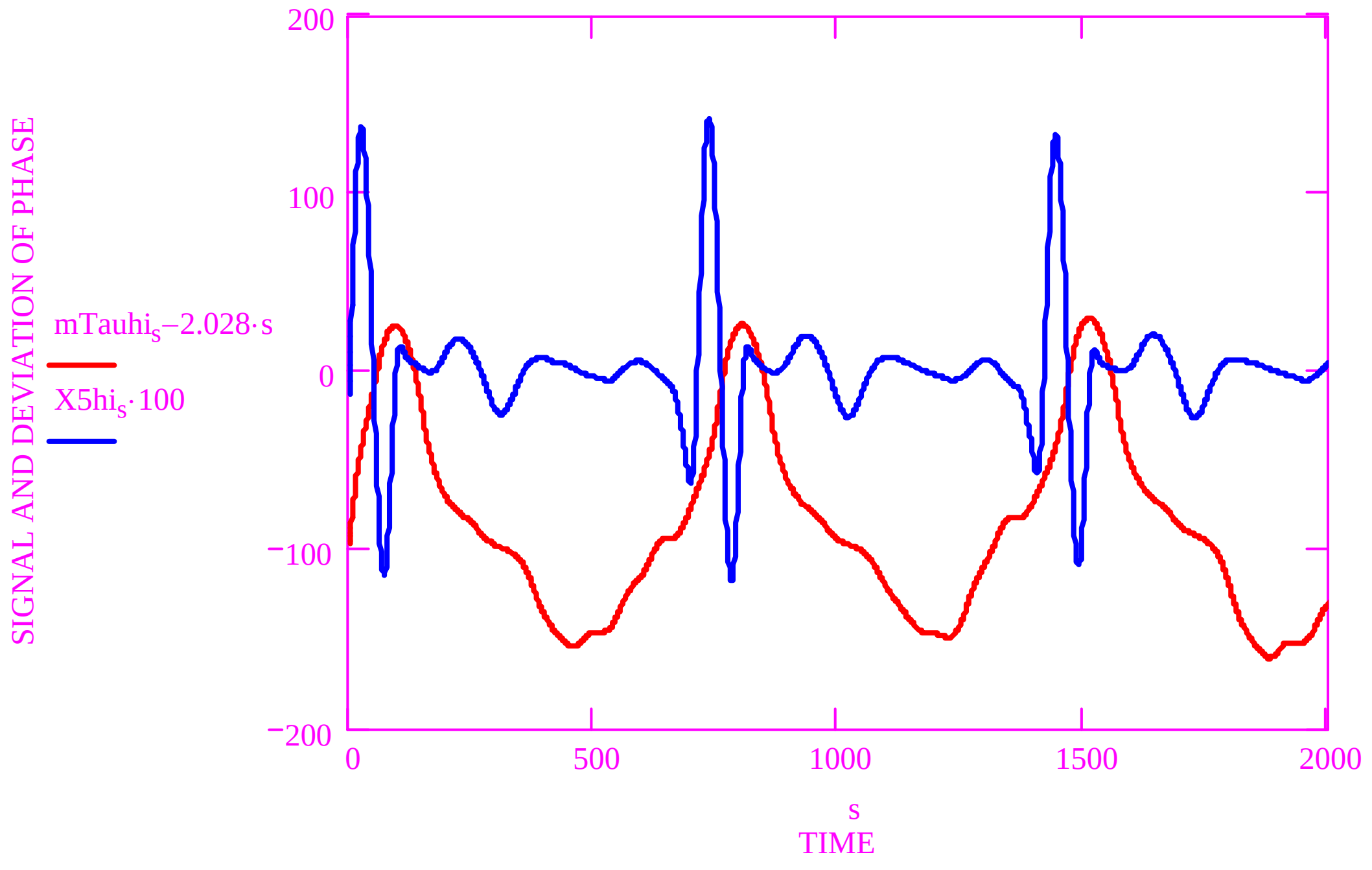

**Derivation of the phase over a long term trend (smoothened) Vs Signal**

## All TSCs are not very near unity

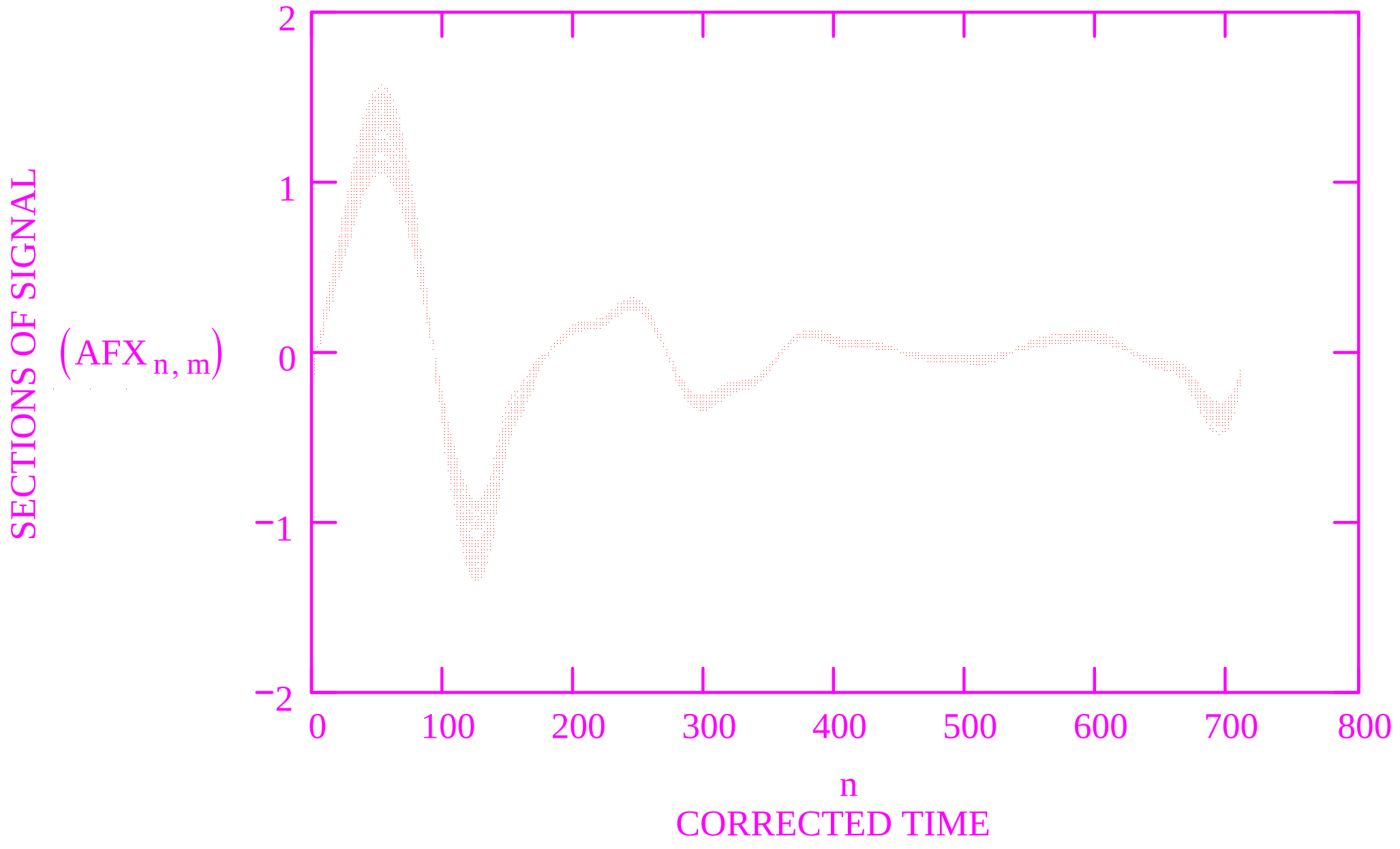

## Rectified signal segments

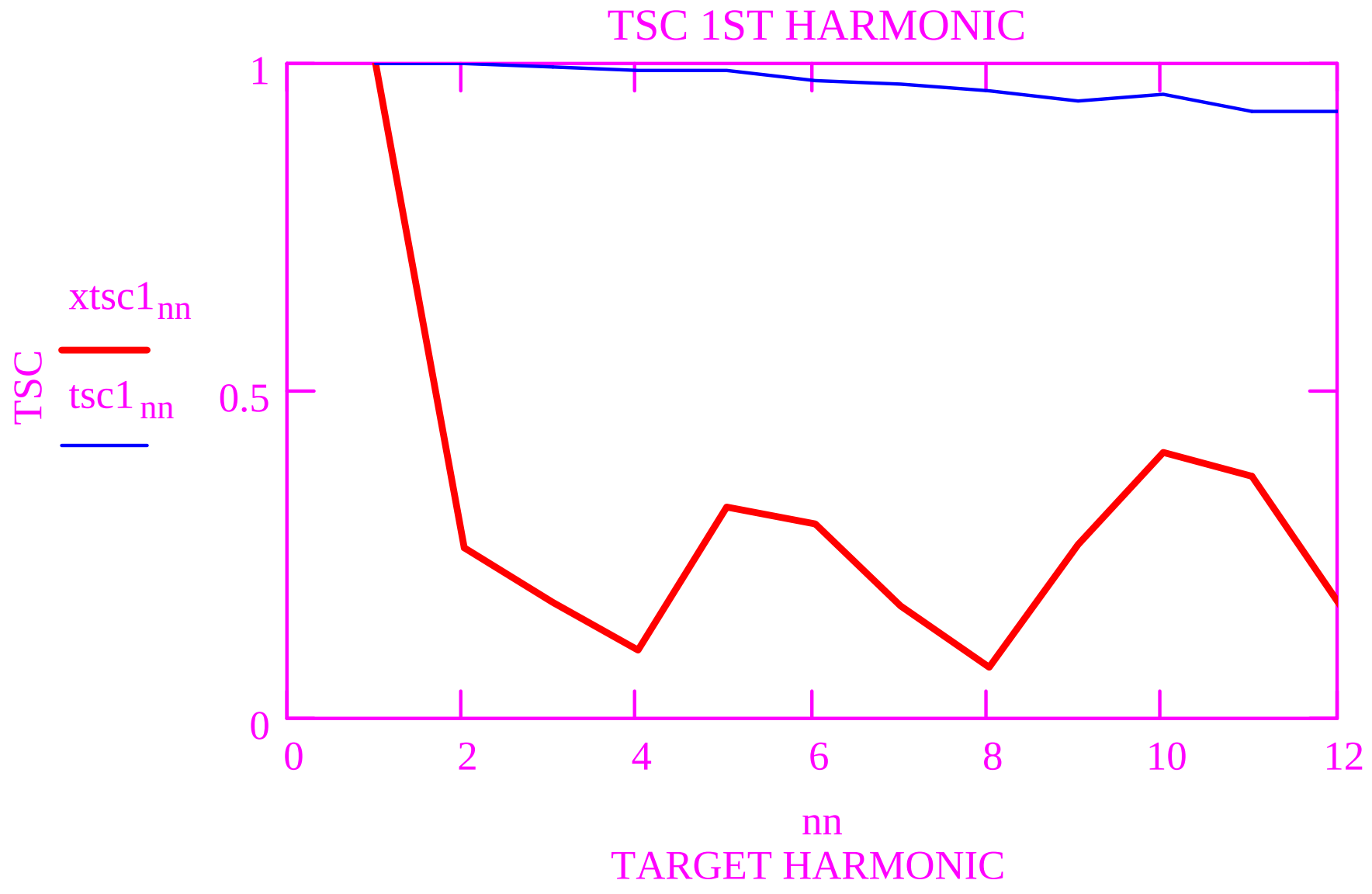

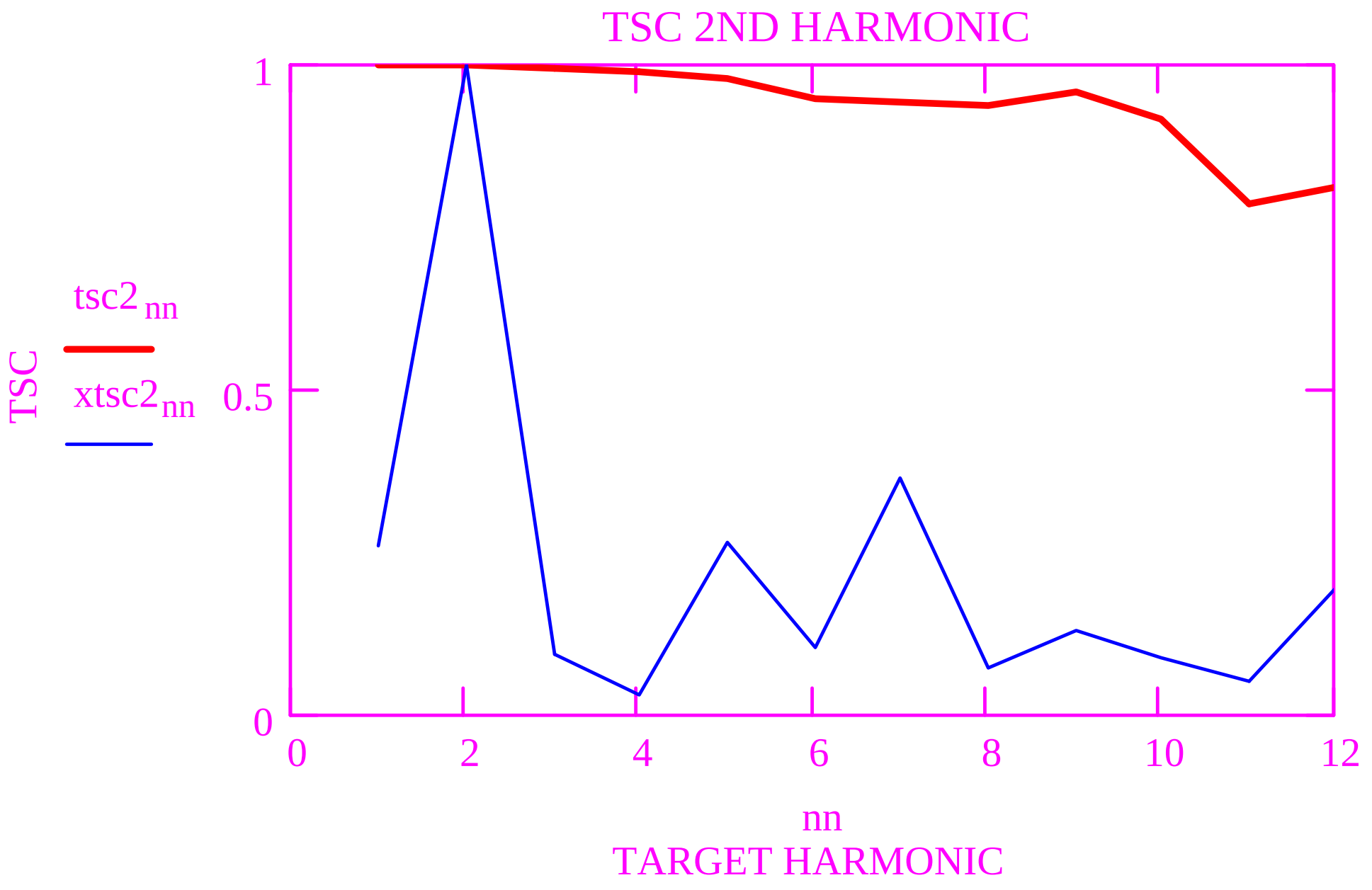

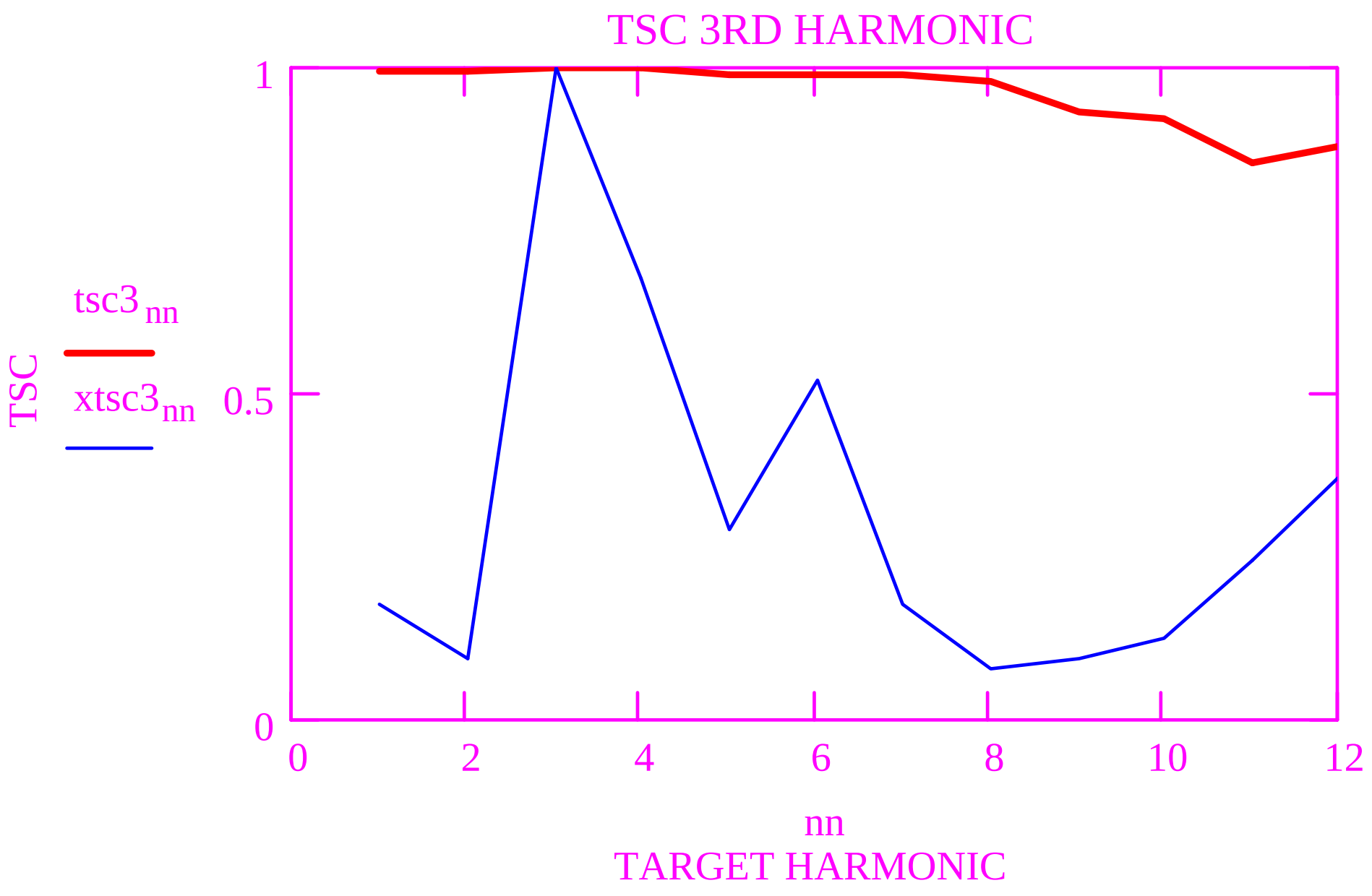

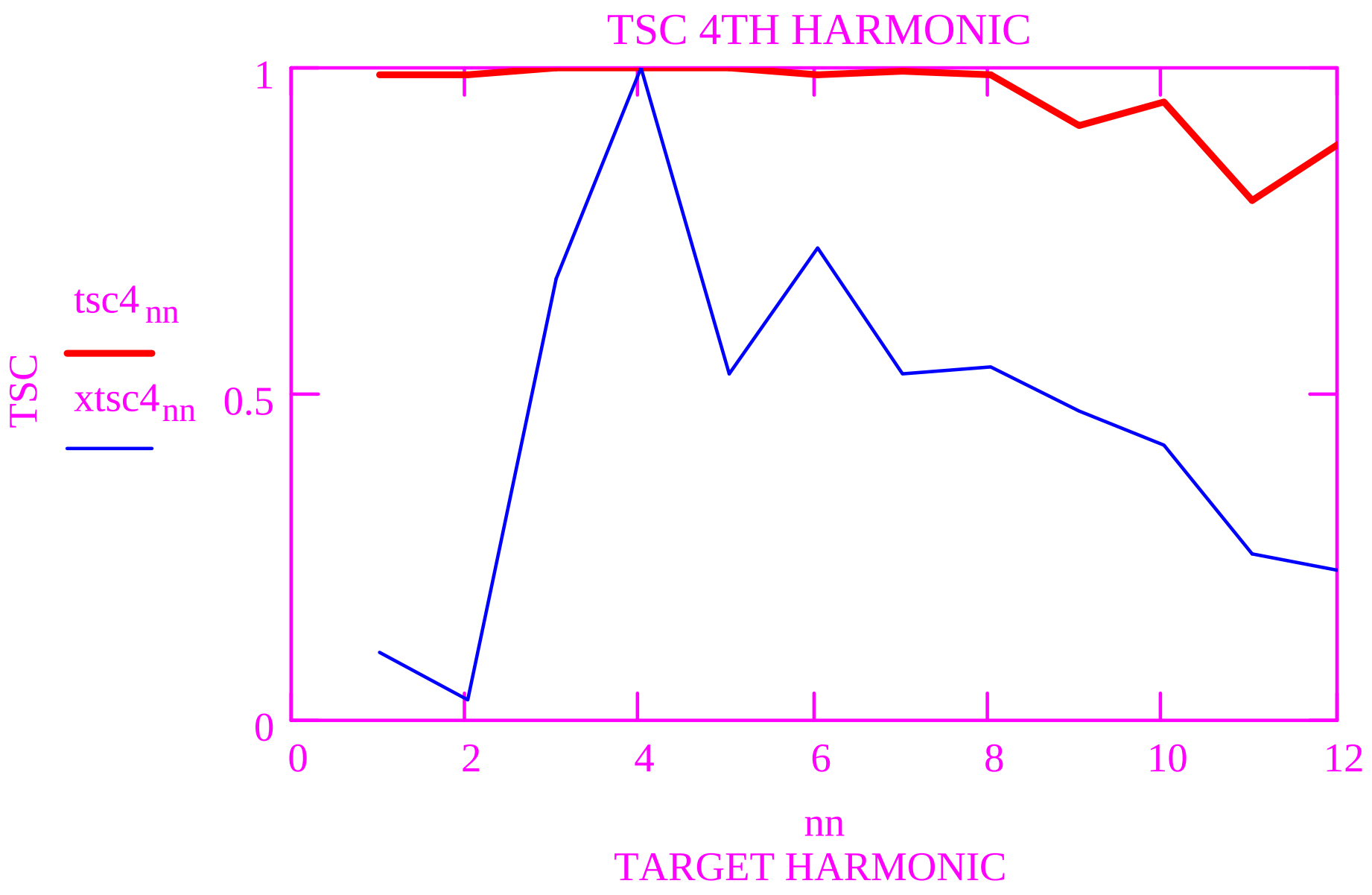

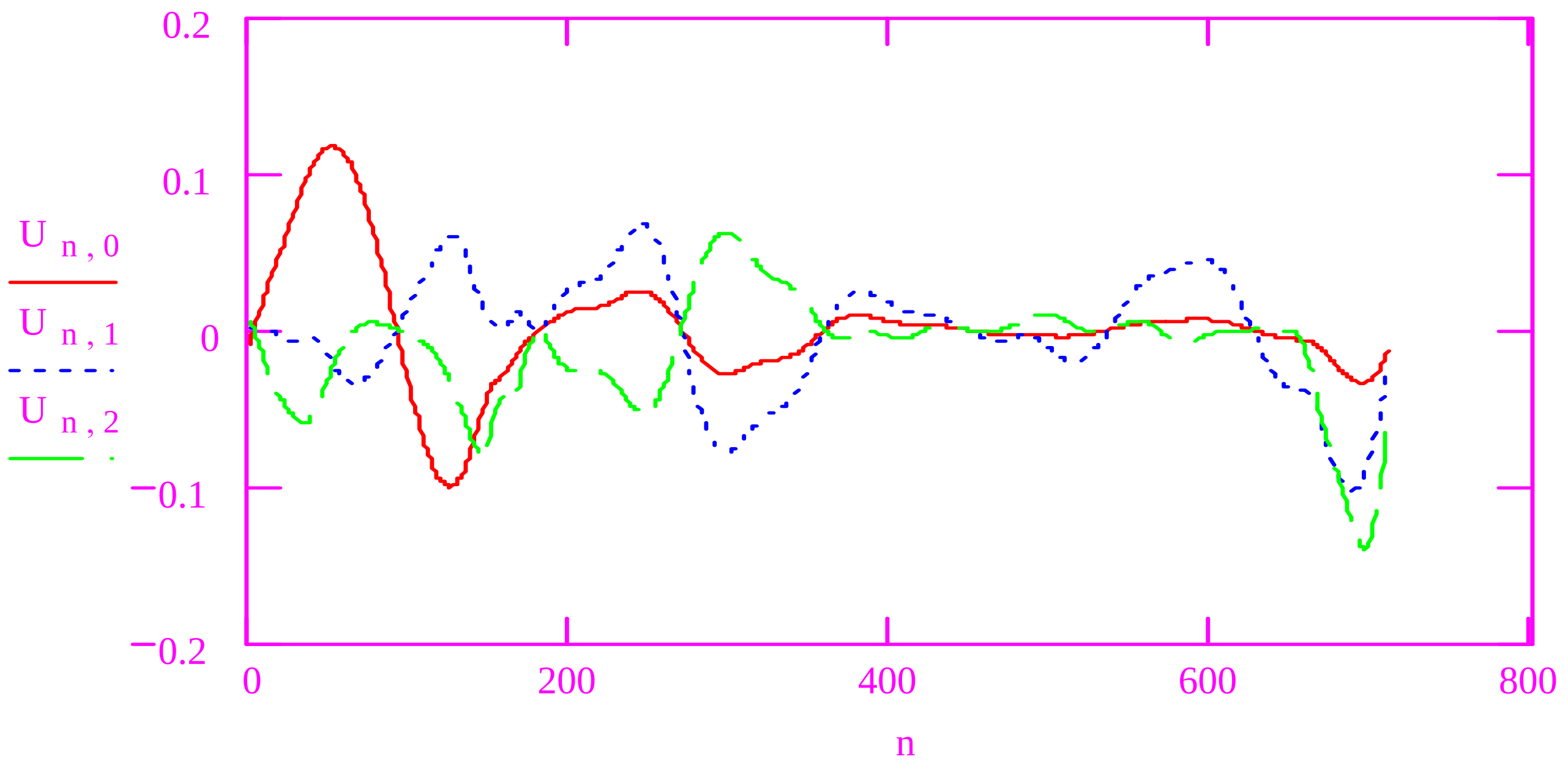

Given the high TSC we look for the evidence of an affine structure in function space.

These are the modes of this structure.

SVS =

| | 0 |
|---|---|
| 0 | 57.87623 |
| 1 | 1.54875 |
| 2 | 1.04607 |
| 3 | 0.56611 |
| 4 | 0.55306 |
| 5 | 0.36452 |
| 6 | 0.31773 |
| 7 | 0.25135 |
| 8 | 0.18777 |
| 9 | 0.15596 |
| 10 | 0.13365 |
| 11 | 0.1034 |
| 12 | 0.08889 |
| 13 | 0.07745 |
| 14 | 0.06486 |
| 15 | 0.05819 |

The time histories of each of the modified windows were used to form columns of a matrix, whose singular values are given in the table.

The singular values show that the most dominant mode is the zeroth mode. The rapid drop in singular values indicate the presence of a low dimensional affine structure.

## Discussion and Conclusion

**We found that using a suitable time transformation, the “phase-resetting” influence on the ECG data can be quite a bit suppressed. The remaining portion has a very simple affine structure. This fact can be utilized in a prediction scheme. (Please see accompanying poster).**